\documentclass{llncs}     

\usepackage[T1]{fontenc}
\usepackage{amsmath}
\usepackage{graphicx}

\begin{document}

\begin{frontmatter}    

\title{Influence of group characteristics on agent voting}

\author{Marcin Maleszka}
\institute{Faculty of Computer Science and Management\\Wroclaw University of Science and Technology\\Wyb. Wyspianskiego 27, 50-370 Wroclaw}
\maketitle
\email{marcin.maleszka@pwr.edu.pl}
\begin{abstract}
A collective of identical agents in a multi-agent system often works together towards the common goal. In situations where no supervisor agents are present to make decisions for the group, these agents must achieve some consensus via negotiations and other types of communications. We have previously shown that the structure of the group and the priority of communication has a high influence on the group decision if consensus theory methods are used. In this paper, we explore the influence of preferential communication channels in asynchronous group communication in situations, where majority vote and dominant value are used. We also show how this relates to consensus approach in such groups and how to use a combination of both approaches to improve performance of real-life multi-agent systems.
\end{abstract}


\end{frontmatter}

\section{Introduction}\label{intro}

Knowledge\footnote{The final publication is available at Springer via https://doi.org/10.1007/978-3-319-54472-4\_6} management and different related tasks are becoming more and more important in modern information society. Different methods of decision making, information retrieval and knowledge integration are being used in various applications, often requiring no input from the user. Over the years, multiple methods have been developed to solve the problem that sometimes occur during these tasks, including consensus theory to handle one cause of the problems in knowledge integration, that is knowledge inconsistency. Inconsistency is a feature of knowledge which may be characterized by the lack of possibility for inference processes, therefore solving it is an essential task in many cases of knowledge management \cite{consensus}. In our overall research we focus on time-related aspects of this problem and in this paper we focus on applying the lessons learned in decentralized multi-agent systems using consensus theory to similar systems using choice theory \cite{choice}.

In our previous papers \cite{aciids2016,smc2016} we have focused mainly on various approaches to collective knowledge integration based on consensus theory \cite{consensus} and tested them both in simulation environment and a prototype practical application. We used asynchronous communication and preferred communication channels to better represent several real-world situations. We researched some previously unexplored aspects of agent knowledge state change when using this type of integration. Here we focus on the area that was already researched more thoroughly by others (e.g. \cite{choice}). As our first runs of the prototype have shown, a using dominant value of the group as a result of integration may give better results in this type of applications. In this paper we make some observations of agent collective behaviour when this method of knowledge integration is used, exploring it both in simulated environment and in real world application.

This paper is organized as follows: in section \ref{rw} we provide a detailed description of research most relevant to ours, mostly focusing on currently described research, but also including our previous work, upon which we based this paper; section \ref{sim} presents a short overview of a multi-agent system with decentralized voting used in a simulated environment, as well as our observations of its various runs; section \ref{forecast} describes a prototype weather prediction system using results from the simulated environment and built upon our previous iterations of this practical application; in section \ref{con} we provide some concluding results, detailing possible further applications of our research.

\section{Related Works}\label{rw}

In our overall research we have considered multiple areas, including multi-agent systems, collective knowledge integration, decentralized systems, asynchronous communication protocols and more. In this section we will provide short overview of various papers most relevant to the research presented in this paper. As far as we know, some parts of our previous research and methodology have not been researched yet. In this paper however, we use methods mostly known in literature in a similar way as our previous approach, leading to some new and interesting results.

Our research stems from the consensus theory and the possibility of consensus changing over time. This is somewhat similar to continuous-time consensus in multi-agent systems, e.g. autonomous robots or network systems \cite{ren}, where it is used for attitude alignment, flocking, formation control, negotiations, etc. One important aspect is finite-time consensus -- if agents will reach consensus in finite time. Another is the stability of multi-agent systems \cite{bhat} -- if all agents knowledge states converge to the same value. This was considered both in centralized and decentralized agent systems \cite{shihua}. Somewhat similar situation is determining the optimal number of experts in group decision making \cite{vandu1} for example by determining when knowledge added by a new agent is lower than some threshold.


Another aspect of our research is the centralized and decentralized approach to communication in multi-agent systems. There are the traffic control systems \cite{traffic1}, where each basic agent controls a single subsystem or functionality (e.g. one crossroads) and the supervisor agent supervises the traffic flow over the whole area. Similarly in decision support systems using multi-agent approach basic agents may communicate between themselves, sharing some basic data, but the overall decision is made by some main facilitator agent \cite{power_rest}. There are also fully decentralized multi-agent systems with no observing agents, e.q. in \cite{choice} the authors show that a group of diverse random agents deciding by majority vote gives better results in the game of Go that a set of uniform agents. A decentralized system may be also used for surveillance \cite{spie}, where each agent has its own knowledge but shares it to influence other agents.

Our approach to decentralized multi-agent system is based on asynchronous communication and some additional structure similar to a social network of agents, which we use for determining additional, preferential communication channels. With possibility of using supervisor agents, we were able to explore both centralized and decentralized systems, e.g. employees in a company and social networks with \textit{friend} relation. We base this on research done in \cite{symfonia}, where it was proposed as a method for improving the teaching process and other knowledge dissemination cases. A similar approach was used in \cite{strong_ties} to show that only strong ties in a social network lead to improvement in groupwork results.

In this paper we use the same approach, but substitute consensus theory \cite{consensus} by voting rules. A similar approach may be found in \cite{choice}, but there the authors work with a small group of agents (up to 25) with diverse knowledge states. The model proposed by authors is universal for finite systems and was tested on a discretized Go game with both uniform and diverse agents. As the theoretical background of this approach is thoroughly explained in literature, we will not be redefining it for purposes of this paper.

\section{Decentralized Multi-agent System in a Simulated Environment}\label{sim}

In our overall research we are using a previously developed simulation environment based on JADE agent framework \cite{jade}. We previously used it to simulate increasing groups of agents trying to achieve consensus \cite{aciids2015} and to observe the influence of the mode of communication on the process of knowledge integration \cite{aciids2016,smc2016}. The centralized nature of JADE framework is used for the purpose of observing the behaviour of a collective of implemented agents and is independent of the communicating group. The collective itself may communicate either in centralized or decentralized manner. As centralized voting was studied multiple times by other researchers, we focused on the decentralized approach.

By the decentralized dominant value voting we understand the following, expanded from \cite{smc2016}:
\begin{itemize}
	\item Each  agent has a number of preferred other agents (\textit{friends}).
	\item Each  agent starts communication in irregular intervals, sending its knowledge to a random other agent, with preference for its \textit{friends}.
	\item Agent communication is unidirectional - after agent sends a message, he does not expect a reply.
	\item Receiver agent gathers incoming knowledge. Once a given number of other agents has contacted it with their knowledge, it integrates it and changes his own knowledge state. The voting approach to integration is based on determining the most often occurring knowledge state and changing own knowledge to this one in the next step.
\end{itemize}

In the framework we use identical \textit{SocialAgents} representing members of the collective and a supervisor gathering the observations, tied to the centralized architecture of JADE. The collective itself operates in a decentralized manner. On initialization the social agents generate some random knowledge (a basic example: a single integer values between $0$ and $k$). In irregular intervals (each tick $T$ each agent has a random chance to initiate communication) these agents will then communicate with each other, sending their own current knowledge state to a random other agent (with probability $p$ to an agent from the whole population, chosen with uniform distribution; and with probability $1-p$ to a single agent from a subgroup of \textit{preferred} agents, chosen with uniform distribution). The receiving agent gathers knowledge from several others and integrates it, changing its own knowledge state. Meanwhile, the centralized observer gathers current knowledge states from all agents and prepares information about the collective as a whole (e.g. it may show the result of a centralized integration from the whole group of agents).

In our previous research we used consensus theory \cite{consensus} as a basis for integration, but our practical application has also shown interesting results when using dominant value \cite{kie}. Following this, we use dominant value as a basis for integration in this series of experiments. Each agent gathers up to $v$ knowledge states from himself and other agents, then changes its own knowledge state to the dominant value of this set. The preferred agent votes are not weighted differently, but their is a higher chance of their knowledge state being included in the voting set.

\begin{figure}[t]
\centering
\includegraphics[width=2.5in]{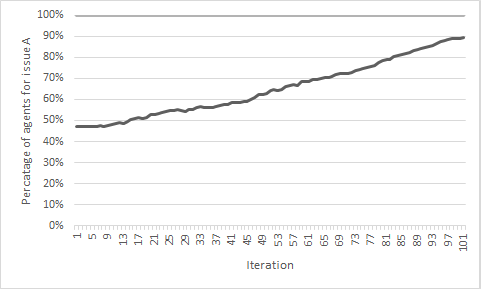}
\caption{Example simulation run for gathering 3 votes and no preferred communications.}\label{fig1}
\end{figure}

\begin{figure}[t]
\centering
\includegraphics[width=2.5in]{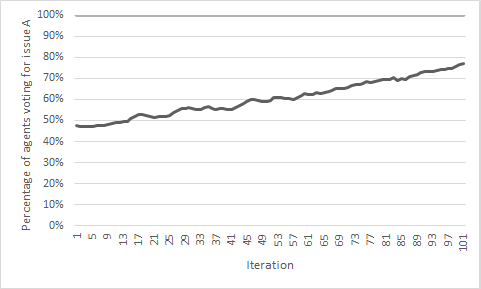}
\caption{Example simulation run for gathering 3 votes, 20 preferred agents and 40\% chance of communication to preferred agents.}\label{fig2}
\end{figure}

\begin{figure}[t]
\centering
\includegraphics[width=2.5in]{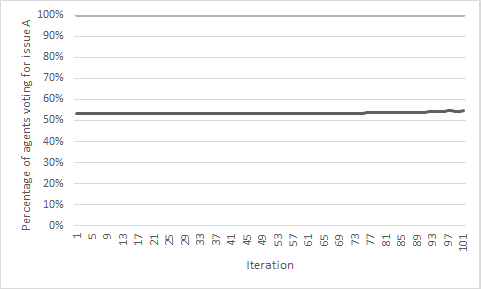}
\caption{Example simulation run for gathering 20 votes, 10 preferred agents and 20\% chance of communication to preferred agents.}\label{fig3}
\end{figure}

We conducted several series of experiments, changing parameters such as type of knowledge, total number of agents, number of preferred agents, probability of communication to preferred agents, number of votes to gather before integration and others. The key observations we gathered may be distilled to the case of a binary choice made by a large group of agents (here: $500$). Some interesting runs are presented in figures \ref{fig1}, \ref{fig2}, \ref{fig3}. We show the \textit{winning} decision as the increasing one in the chart.


Our observations of the simulation runs may be outlined as follows:

The smaller the number of different opinions gathered, the higher the (long term) chance that the agent will change its own knowledge state. This corresponds to earlier research in \cite{choice} and \cite{strong_ties}. The first paper shows, that for small number of diverse agents in the group, the results are more often biased towards one side (in case of the game of Go this leads to losing strategy). The latter describes the case of only strong ties in a social network actually influencing the groupwork -- here this corresponds to stronger influence of a single relation if the number of relations (votes) is smaller.

For larger number of gathered outside opinions, the chance of the agent changing its knowledge state is significantly smaller, as the opinions will correspond more closely to the division of knowledge in the overall group.

The larger the chance of selecting preferential communication channels, the higher the chance that the agents in \textit{friend} relation will influence the opinion of an agent (ie. the agents opinion will be closer to the subgroup opinion). This leads to certain subgroups of the collective becoming more distinct from others, but uniform among their members.

On the other hand, a large number of diverse agents in the preferred group lowers the chance of the agent changing its opinion. As a large group reflects the whole collective more closely than a smaller one, the chance that the group is biased towards some opinion is lower. This observation is especially important for our practical application, as it allows fine tuning of the integration process -- the number of gathered votes determines the overall trend and the size of \textit{friend} group regulates the speed of the occurring change.

\section{Application: Weather Forecast System}\label{forecast}

During our research of the influence of agent communication on knowledge integration, we are using a prototype of a weather forecasting multi-agent system using our theoretical and simulation results. In our early research \cite{kie} we have observed that using dominant value in integration, instead of consensus theory approach, gave slightly better results. This spurned the research presented in this paper. After conducting the simulation runs, we were able to further fine tune our prototype application and improve the predictions of the whole system.

The architecture of the prototype system consists of three distinct layers:
\begin{itemize}
	\item Source layer - a collective of various agents (corresponding to \textit{SocialAgents} in the simulation environment) with two functions: creating their own predictions (based on historical data, copied from internet sources, etc.) and using the decentralized voting approach to modify their predictions based on other agents forecasts. Different integration methods may be used for modifying agents knowledge, both based on consensus theory and dominant value. Some agents may drop out when their sources are unavailable, but this does not influence overall system.
	\item Integration layer - after the time for individual communication between agents ends, a single supervisor agent gathers results from all agents and integrates them to a single forecast, using different approaches (consensus or dominant value).
	\item Interaction layer - in a practical system this is the interface layer, presenting the final forecast to the user. In experimental setup this is the observation agent, similar to the one used in the simulation environment. It gathers both the final forecast and any partial results and preprocesses them for later analysis.
\end{itemize}

We are conducting evaluations of the weather prediction system using real world data in Wroclaw, Poland. For each iteration of the system (with added different integration methods), we have conducted experiments on all its variants (basic dominant value and centralized consensus in April-May 2015, centralized and decentralized consensus in October 2015, centralized consensus, decentralized consensus and decentralized consensus with \textit{friend} relation in April-May 2016). Based on simulation environment results, we decided to test various methods of integration in source layer combined with using dominant value for integration layer. As previously we have run the other variants of the weather prediction system in parallel, to compare their effectiveness. The results for all time periods and methods are shown in Table \ref{tab_pract}.

\begin{table}
\caption{All observed runs of the weather prediction system in all variants: basic dominant value (B. Dominant), centralized consensus (Cen. Cons.), decentralized consensus (Dec. Cons.), decentralized consensus with \textit{friend} relation (D.-S. Cons.) and two new variants of dominant value approach: full decentralized voting in source layer (Dom. Dec.) and voting interchangeable with consensus in source layer (Dom. Mix.)}
\label{tab_pract}
\centering
\begin{tabular}{|c|c|c|c|c|}
  \hline 
  System-Run & MAE & Comp. w/ & Comp. w/ & Comp. w/ \\
	& & Best Src. & Worst Src. & Avg. Src. \\
  \hline
  \hline
	B. Dominant (IV-V '15) & 1,857 & 89\% & 17\% better & 3\% better \\
	\hline
	Cen. Cons. (IV-V '15) & $2,018$ & 82\% & 7\% better & 95\% \\
	\hline
	\hline
  Cen. Cons. (X '15) & $2,132$ & 75\% & 2\% better & 90\% \\
	\hline
	Dec. Cons. (X '15) & $1,984$ & 83\% & 9\% better & 97\% \\
	\hline	
	\hline
	Cen. Cons. (IV '16) & $1,991$ & 85\% & 6\% better & 93\% \\
	\hline
	Dec. Cons. (IV '16) & $1,994$ & 85\% & 6\% better & 93\% \\
	\hline
	D.-S. Cons.(IV-V '16) & $1,989$ & 85\% & 6\% better & 93\% \\
	\hline
	\hline
	B. Dominant (X '16) & 1,193 & 89\% & 15\% better & 99\% \\
	\hline
	Cen. Cons. (X '16) & 1,956 & 87\% & 12\% better & 97\% \\
	\hline
	Dec. Cons. (X '16) & 1,931 & 88\% & 14\% better & 98\% \\
	\hline
	D.-S. Cons. (X '16) & 1,933 & 88\% & 14\% better & 98\% \\
	\hline
	Dom. Dec. (X '16) & 1,898 & 90\% & 16\% better & \textit{equal} \\
	\hline
	Dom. Mix. (X '16) & 1,892 & 90\% & 16\% better & 1\% better \\
	\hline
\end{tabular}
\end{table}

Our previous research focused on the consensus based approach to integrate knowledge, but for this practical application repeated experiments have shown that various approaches based on dominant value give better results. The three tested variants of dominant value approach may be described as follows:
\begin{itemize}
	\item Basic Dominant Value -- this most basic approach does not require any additional communication in the source layer. The basic agents calculate their own forecast and the supervisor agent in the integration layer selects the most often occurring value as the final prediction.
	\item Dominant Value in Decentralized System -- in this approach the agents in the source layer have some time to communicate and change their predictions based on other agents information, before the supervisor agent selects the most often occurring value as the final prediction. The source layer agents communicate as in the simulation environment with tuned parameters: gathering 3 votes, 15 preferred agents and 40\% chance of communication to preferred agents.
	\item Dominant Value mixed with Consensus Integration -- this approach is similar to the second one, but there is additional 50\% chance that the knowledge change in the source layer will be based on consensus theory (best results as presented in \cite{smc2016}) instead of dominant value.
\end{itemize}

Our overall experiments with the prototype weather prediction system have shown that the best results were determined using the Dominant Value mixed with Consensus Integration approach. In each tested situation the various dominant value approaches had the smallest MAE and were the closest to the single most accurate prediction.

\section{Conclusions}\label{con}

This paper finalizes our research into observing the changes of knowledge states of agents in a decentralized collective during integration process. Previously we have studied centralized systems and various types of decentralized systems with integration accomplished by means of consensus theory. In this paper we used the voting methods described in literature and applied them to the same type of system. We have observed the behaviour of such system in a simulated environment, gathering some guidelines for fine tuning the integration process. We then applied these observations to a prototype weather prediction system we have been developing parallel to our research. Here we have shown that a specific mix of voting and consensus, used in a hybrid centralized-decentralized system provides best results for this specific practical applications.

Our future research also includes developing our own agent platform, independent on frameworks such as JADE. We will use it both as a basis for  the next version of our simulation environment and for future version of the prototype weather forecast application. Disassociating our research from platform limitations should allow for larger scale experiments and platform independent applications. This will allow us to implement our other prototype applications, such as economic and traffic prediction systems. In particular for a traffic system we consider single agents observing single drivers in different road situations and monitoring them continuously. After some finite time a single agent should determine if the driver is a good or a bad one (this is more related to our research in \cite{aciids2015},
which would especially important to insurance companies. Additionally, monitoring individual drivers on a local (via decentralized integration) and global (centralized) level would allow improvements to the traffic flow. We also consider the same approach to determine important, but not explicit topics discussed in web commentaries.

\section*{Acknowledgment}

This research was co-financed by Polish Ministry of Science and Higher Education grant.



\begin{thebibliography}{9}
			
  \bibitem{bhat} Bhat S.P., Bernstein D.S.: Finite-time Stability of Continuous Autonomous Systems. In: Siam J. Control Optim. 38 (3),  751--766 (2000)
  
  \bibitem{strong_ties} De Montjoye Y.-A., Stopczynski A., Shmueli E., Pentland A., Lehmann S.:The Strength of the Strongest Ties in Collaborative Problem Solving. Scientific reports 4, Nature Publishing Group 2014.
	
		\bibitem{traffic1} Iscaro G., Nakamiti G.: A supervisor agent for urban traffic monitoring. IEEE International Multi-Disciplinary Conference on Cognitive Methods in Situation Awareness and Decision Support (CogSIMA), IEEE 2013, pp. 167--170. 
	
	  \bibitem{jade} JADE, Java Agent Development Framework, available at: http://jade.tilab.com/ 
	
		\bibitem{choice} Jiang A., Marcolino L. S., Procaccia A. D., Sandholm T., Shah N., Tambe M.: Diverse randomized agents vote to win. Advances in Neural Information Processing Systems, 2014, pp. 2573--2581.
  	\bibitem{shihua}  Li S., Dua H., Lin X.: Finite-time consensus algorithm for multi-agent systems with double-integrator dynamics. In: Automatica 47, 1706--1712 (2011)
		
					\bibitem{aciids2015} Maleszka M.: Consensus with Expanding Conflict Profile. New Trends in Intelligent Information and Database Systems
Studies in Computational Intelligence Vol. 598, Springer, 2015 pp. 291--299
			
			\bibitem{aciids2016} Maleszka M.: Knowledge in Asynchronous Social Group Communication. Intelligent Information and Database Systems,Lecture Notes in Computer Science Vol. 9621, Springer, 2016, pp. 364--373.
			
				\bibitem{symfonia} Maleszka M., Nguyen N.T., Urbanek A., Wawrzak-Chodaczek M.: Building Educational and Marketing Models of Diffusion in Knowledge and Opinion Transmission. Computational Collective Intelligence, Technologies and Applications, Lecture Notes in Artificial Intelligence, Springer International Publishing, 2014, pp. 164--174
				
					\bibitem{smc2016} Maleszka M.: Local and Global Consensus in Asynchronous Group Communication. Proceedings of IEEE SMC 2016 (in print)
			
		\bibitem{kie} Mercik J., Tolkacz O., Wojciechowska J., Maleszka M.: Wykorzystanie integracji wiedzy do zwiekszenia efektywnosci prognozowania w warunkach niepewnosci. Porebska-Miac T. (Ed.), Systemy Wspomagania Organizacji 2015, Wydawnictwo Uniwersytetu Ekonomicznego w Katowicach, Katowice, 2015.

  			\bibitem{consensus} Nguyen N.T.: Advanced methods for inconsistent knowledge management. Springer, 2007.
				
					\bibitem{power_rest} Nagata T.,  Sasaki H.: A multi-agent approach to power system restoration. IEEE Transactions on Power Systems Vol.17(2), IEEE 2002, pp. 457--462.

	\bibitem{vandu1} Nguyen V.D., Nguyen N.T.: An Influence Analysis of the Inconsistency Degree on the Quality of Collective Knowledge for Objective Case. Intelligent Information and Database Systems, Lecture Notes in Computer Science Vol. 9621, Springer, 23--32, 2016
	
	\bibitem{spie} Peterson C. K., Newman A. J., Spall J. C.: Simulation-based examination of the limits of performance for decentralized multi-agent surveillance and tracking of undersea targets. SPIE Defense+ Security, International Society for Optics and Photonics, 2014, pp. 90910F--90910F.
		
	
\bibitem{ren} Ren W., Beard R.W., Atkins E.M.: A Survey of Consensus Problems in Multi-agent Coordination. American Control Conference, 2005. Proceedings of the 2005. IEEE, 1859--1864 (2005) 
	


\end{thebibliography}
\end{document}